\documentclass[11pt]{article}
\usepackage{graphicx}
\usepackage{amssymb}
\renewcommand{\baselinestretch}{2}

\begin{document}
\title{\bf  Model reconstructions for the Si(337) orientation}
\author{Feng-Chuan Chuang\\
Ames Laboratory -- US Department of Energy, and \\Department of
Physics, Iowa State University, Ames, IA 50011, USA \\
Cristian V. Ciobanu\footnote{Corresponding author, email:
cciobanu@mines.edu}\\Division of Engineering, Colorado School of
Mines, Golden, CO 80401, USA \\
Cai-Zhuang Wang and Kai-Ming Ho\\
Ames Laboratory -- US Department of Energy, and \\Department of
Physics, Iowa State University, Ames, IA 50011, USA}
\maketitle  %
\begin{abstract}
{Although unstable, the Si(337) orientation has been known to
appear in diverse experimental situations such as the nanoscale
faceting of Si(112), or in the case of miscutting a Si(113)
surface. Various models for Si(337) have been proposed over time,
which motivates a comprehensive study of the structure of this
orientation. Such a study is undertaken in this article, where we
report the results of a genetic algorithm optimization of the
Si(337)-$(2\times 1)$ surface. The algorithm is coupled with a
highly optimized empirical potential for silicon, which is used as
an efficient way to build a set of possible Si(337) models; these
structures are subsequently relaxed at the level of ab initio
density functional methods. Using this procedure, we retrieve the
(337) reconstructions proposed in previous works, as well as a
number of different ones. }
\end{abstract}
\newpage

\section{Introduction}

Silicon surfaces have been at the foundation of the semiconductor
industry for several decades. With the advent of nanotechnology,
the interest in the atomic structure of various semiconductor
surfaces has expanded because of their potential to serve as
natural and inexpensive templates for growing various types of
nanostructures. Understanding the growth of nanoscale entities on
various substrates depends, at least in part, on knowing the
atomic structure of the substrate. As the low-index crystal
surfaces are well understood, the high-index orientations are now
progressively gaining importance as their richer morphological
features may lead to more advanced technological applications. To
date, there are numerous stable high-index orientations reported
in the literature; in the case of Silicon and Germanium, these
stable orientations have been summarized in
Ref.~\cite{Si-high-index-Yang}. The {\em unstable} high-index
orientations are important in their own right, as they often give
rise to remarkable periodic grooved morphologies which may be used
for the growth of surface nanowires. The wonderful morphological
and structural diversity of high-index surfaces can be
appreciated, for instance, from the work of Baski, Erwin and
Whitman, who investigated systematically the surface orientations
between Si(001) and Si(111) \cite{001-111-baski}.

An interesting case of unstable surface is Si(337), whose atomic
structure is the subject of the present article. As a historical
account, we note that the Si(337) orientation was concluded to be
stable in early studies by Ranke and Xing \cite{ranke-337-stable},
Gardeniers {\em et al.} \cite{337Gardeniers1989}, and Hu {\em et
al.} \cite{337-applSSci}. This conclusion about the stability of
Si(337) was consistent with the reports of Baski and Whitman, who
showed that Si(112) decomposes into quasi-periodic Si(337) and
Si(111) nanofacets \cite{112}, and that Si(337) is a low-energy
orientation \cite{112-H}. However, further studies by Baski and
Whitman revealed that Si(337) itself facets into Si(5 5 12) and
Si(111) \cite{Ga-induced-restruct-112-337}, demonstrating that
Si(337) was, in fact, unstable. Since Si(337) is not stable and
Si(5 5 12) is \cite{5512science, 5512Mochrie}, the clean Si(112)
orientation should facet into Si(5 5 12) and Si(111). However, as
explained in Ref. \cite{Ga-induced-restruct-112-337}, nanofacets
are too narrow to form complete Si(5 5 12) units, and mostly
settle for the nearby Si(337) orientation. A more recent report
\cite{337step-Takeguchi} shows another situation where Si(337)
arises rather than the expected Si(5 5 12). For a Si(113) surface
that is miscut towards [111], the large (5 5 12) reconstruction
has not always been found. Instead, a Si(337) phase has been seen
to coexist with Si(113) terraces \cite{337step-Takeguchi}. The
high resolution transmission electron microscope (HRTEM) images in
Ref. \cite{337step-Takeguchi} leave no doubt that the periodicity
of those nanofacets corresponds to the (337) orientation. The
reason for the quasi-stability of Si(337) reported in earlier
works \cite{ranke-337-stable, {337Gardeniers1989}, 337-applSSci}
could be the curvature or the size of substrates used in those
investigations; this explanation is also consistent with Ref.
\cite{337step-Takeguchi}, where  the issue of sample curvature is
acknowledged.

With one exception \cite{337-applSSci}, atomic-scale models for
Si(337) were not sought separately, but rather as structural parts
of the Si(5 5 12)-($2\times 1$) reconstruction
\cite{5512science,5512ranke,{5512-TEM-takeguchi},
{5512-Jeong-yr2004}}. As shown in Ref.~\cite{5512science}, the
Si(5 5 12) unit cell consists of two Si(337) units and one Si(225)
unit. The first model reconstruction proposed for Si(5 5 12)
\cite{5512science} appeared somewhat corrugated when viewed along
the [${\overline 1}10$] direction; so did a second model by Ranke
and Xing \cite{5512ranke}. On the other hand, HRTEM measurements
\cite{5512-TEM-takeguchi} showed a flatter profile for Si(5 5 12),
and different model reconstructions were proposed
\cite{5512-TEM-takeguchi, {5512-Jeong-yr2004}} in order to account
for the observed flatness. Total-energy density functional
calculations for the energy of Si(5 5 12) have only  recently been
published, and suggest that the latest models
\cite{5512-TEM-takeguchi, {5512-Jeong-yr2004}} have lower energies
than the earlier proposals \cite{5512science, 5512ranke}.

The reason structural studies of high-index surfaces can be very
lengthy originates from the way model reconstructions are
proposed,  which is based on physical intuition in interpreting
scanning tunneling microscope (STM) images. As seen above for the
case of Si(5 5 12), the intuitive ways are not very robust for
high-index surfaces, since usually more models are required when
further experiments are reported. Given the rather large number of
low-energy structures that high-index surfaces can have (see,
e.g., the case of Si(105)\cite{105SSLett,ptmc105}), heuristic
approaches pose the risk of missing the actual physical
reconstruction when considering model candidates for comparison
with the experimental data. To alleviate this risk, we have
recently addressed the reconstruction of semiconductor surfaces as
a problem of stochastic optimization \cite{ptmc105}, and developed
two different global search algorithms for the purpose of
determining the most favorable surface structure \cite{ptmc105,
ga105}. We have used these global search procedures for studying
Si(105) \cite{ptmc105,ga105} and Si(114) \cite{ptmc-ga-114}, which
are both stable orientations for silicon. Although Si(337) is not
a stable orientation, its atomic structure can still be pursued
with global search algorithms because the only input required is
the surface periodicity, which is available from STM
\cite{5512science} and LEED \cite{337-applSSci} measurements.

In this article we report the results of such a global
optimization for the structure and energy of Si(337) using
empirical potential and density functional theory calculations. As
we shall see, the optimization retrieves most of the previously
proposed models \cite{337-applSSci,
5512science,5512-TEM-takeguchi,5512-Jeong-yr2004}, as well as a
number of other reconstructions that could be relevant for
experimental situations where Si(337) nanofacet arises ---other
than the two (337) unit cells that appear as part of the (5 5 12)
structure. The remainder of this paper is organized as follows. In
the next section we briefly describe the computational methods
used in this study. In Section 3, we present our model structures
for the Si(337) surface classified by the number of atoms in the
computational cell and their surface energy. In Section 4, we
compare our results with other reported Si(337) reconstructions.
Our conclusions are summarized in the last section.

\section{Methods and Computational details}

The Si(337) simulation cell consists of two (337) bulk truncated
unit cells of dimensions $a\sqrt{8.375} \times a\sqrt{0.5} \times
a\sqrt{67}$, ($a$ is the lattice constant of Si), stacked along
the $y$-direction, $[{\overline 1}10]$. These dimensions are
calculated from the crystal geometry, and are in agreement with
the STM images published in Ref. \cite{5512science}, where (337)
units are shown as part of a stable reconstructed surface, Si(5 5
12)-$(2\times 1)$. Because the (337) plane has two-dimensional
periodicity along two {\em non-orthogonal} directions, the
rectangular unit in Fig.~\ref{supercell} must be subjected to
shifted periodic boundary conditions \cite{farid}, with the shift
in the $y$-direction equal to $a/\sqrt{8}$. We build a database of
structures sorted out by the number of atoms in the simulation
cell and by their surface energy calculated with the highly
optimized empirical potential developed by  Lenosky {\em et al.}
\cite{hoep}. The number of atoms is varied around the initial
value of $n=268$, in order to exhaust all the distinct
possibilities that can occur during the search for low energy
reconstructions.

The global optimization involved in building the structure
database for Si(337) has been performed with the genetic algorithm
adapted for crystal surfaces \cite{ga105}. This algorithm is based
on an evolutionary approach in which the members of a generation
(pool of models for the surface) evolve with the goal of producing
the best specimens, i.e. lowest energy reconstructions. The
evolution from one generation to the next takes place by mating,
which is achieved by subjecting two randomly picked structures
({\em parents}) from the pool  to an operation that combines their
surface features and creates a {\em child} configuration. This
child structure is relaxed and considered for inclusion in the
pool, depending on its surface energy. We use two variants of the
algorithm, one in which the number of atoms is kept constant for
all members of the pool at all times (the constant-$n$ version),
and another one in which this restriction is not enforced (the
variable-$n$ version). The implementation  of this genetic
algorithm has recently been described in various degrees of detail
in Refs. \cite{ga105, ptmc-ga-114, ga-perspective}; for this
reason, we will not expand upon it here, but refer the reader to
those works. Since the empirical potential may not give a very
reliable energy ordering for a database of structures, we study
not only the global minima given by HOEP for different values of
$n$, but also most of the local minima that are within 15
meV/\AA$^2$ from the lowest energy configurations. After creating
the set of Si(337) model structures via the genetic algorithm, we
recalculate the surface energies of these minima at the level of
density functional theory (DFT).

The DFT calculations where performed with the plane-wave based
PWscf package \cite{pwscf}, using the Perdew-Zunger \cite{Perdew}
exchange-correlation energy. The cutoff for the plane-wave energy
was set to 12 Ry, and the irreducible Brillouin zone was sampled
using 4 $k$-points. The equilibrium bulk lattice constant was
determined to be $a=$5.41\AA, which was used for all the surface
calculations in this work. The simulation cell has the single-face
slab geometry with a total silicon thickness of 10\AA \ , and a
vacuum thickness of 12 \AA. The Si atoms within a 3\AA-thick
region at the bottom of the slab are kept fixed in order to
simulate the underlying bulk geometry, and the lowest layer is
passivated with hydrogen. The remaining Si atoms are allowed to
relax until the force components on any atom become smaller than
0.025 eV/\AA.

The surface energy $\gamma$ for each reconstruction is determined
indirectly, by first considering the surface energy $\gamma_0$ of
an unrelaxed bulk truncated slab, then by calculating the
difference $\Delta \gamma = \gamma - \gamma _0 $ between the
surface energy of the actual reconstruction and the surface energy
of a bulk truncated slab that has the bottom three layers fixed
and hydrogenated. This indirect method for calculating the surface
energies at the DFT level was outlined, for instance, in Ref.
\cite{113dabrowski}. The energy of the bulk truncated surface
shown in Fig.~\ref{supercell} was found to be $\gamma_0=137.87$
meV/\AA$^2$. At the end of the genetic algorithm search, we obtain
a set of model structures which we sort by the number of atoms in
the simulation cell and by their surface energy. These model
structures are presented next.

\section{Results}

There are four possibilities in terms of the number of atoms in
the slab that yield distinct global energy minima of the Si(337)
periodic cell shown in Fig.~\ref{supercell}. This has been
determined by performing constant-$n$ genetic algorithm
optimization for computational slabs with consecutive numbers of
atoms $n$ $(264\leq n \leq 272)$, and identifying a periodic
behavior of the lowest surface energy as a function of $n$. This
procedure of determining the symmetry distinct numbers of atoms in
the slab has been detailed in Ref.~\cite{ptmc105}. The HOEP global
minima, as well as selected local minima of the surface energy for
different numbers of atoms are summarized in Table~\ref{table},
along with the density of dangling bonds per unit area and the
surface energy computed at the DFT level.

As a general comment, the DFT energy ordering does not coincide
with that given by HOEP, indicating that the transferability of
HOEP to Si(337) is not as good as in the case of Si(001) and
Si(105); to cope with this transferability issue, we have
considered more local minima (than listed in Table~\ref{table})
when performing DFT relaxations, as mentioned in Section 2. In the
terminology of potential energy surface (PES) theory, we sample
the main basins using the genetic algorithm coupled with HOEP,
then recalculate the energetic ordering of these basins using DFT.
From the table it is apparent that the least favorable number of
atoms is $n=267$ (modulo 4) at both the HOEP and DFT levels.
Therefore, we focus on describing here the reconstructions that
have numbers of atoms $n=266,268$, and 269.

For $n=266$ the best model that we obtained at the DFT level is
made of a fused assembly of dimers, tetramers and slightly
puckered honeycombs, followed (in the $[{77\overline 6}]$
direction) by a row of rebonded atoms. These individual
atomic-scale motifs have been reported previously for different
surfaces (e.g., \cite{5512science, 5512-Jeong-yr2004, 114-erwin}),
and depicted for convenience in Fig.~\ref{features}. Their complex
assembly shown in Fig.~\ref{266-fig}(a) has a surface energy of
94.47 meV/\AA$^2$; the reconstruction \ref{266-fig}(a) has a
corrugated aspect when viewed from the $[{\overline 1}10]$
direction, which may account for its relatively high surface
energy (refer to Table~\ref{table}). Higher-energy models with
$n=266$ are illustrated in Fig.~\ref{266-fig}(b,c,d), and have
surface energies ranging from 2.75 meV/\AA$^2$ to about 7
meV/\AA$^2$ above the surface energy of the model
\ref{266-fig}(a). Dimers (D) and tetramers (T) can be clearly
identified on model \ref{266-fig}(b), while configuration
\ref{266-fig}(c) is characterized by the presence of honeycombs
(H) and rebonded atoms (R). In model \ref{266-fig}(d) the only
features from the set in Fig.~\ref{features} that can be
separately identified are the rebonded atoms R. Nearly degenerate
with structure 3(d), we find two other models [\ref{266-fig}(e)
and \ref{266-fig}(f), with different tilting of the dimers
belonging  to tetramer groups] that contain 6-member ring
$\pi$-chains \cite{5512science}. The 6-member rings are labelled
by 6-r in Fig.~\ref{266-fig}(e,f), and are supported by
tetramer-like features (denoted by t in Fig.~\ref{266-fig}) whose
dimers are made of fully coordinated atoms.

The optimum number of atoms is $n=268$, as indicated in
Table~\ref{table} at both HOEP and DFT levels --coincidentally,
this is also the number of atoms that corresponds to two complete
(337) bulk truncated primitive cells. We have further verified
this number by performing a variable-$n$ genetic algorithm started
with all the members in the pool initially having $n=267$ atoms.
The algorithm has found the correct number of atoms ($n=268$) and
the HOEP global minimum in less than 1000 operations. While
different starting configurations may change this rather fast
evolution towards the global minimum, our tests indicate that for
simulation cells of similar size several thousand genetic
operations are sufficient when using the implementation described
in Ref.~\cite{ga105}. The lowest energy model as given by DFT is
shown in Fig.~\ref{268-fig}(a) and consists of dimers D, rebonded
atoms R and honeycombs H, in this order along $[77{\overline 6}]$.
We note that the number of dangling bonds is smaller at the DFT
level (refer to Table~\ref{table}): the reason for the dangling
bond reduction is the flattening of the honeycombs (not planar at
the level of empirical potentials), which creates two additional
bonds with the subsurface atoms. The genetic algorithm search
retrieves another low-energy model, in which the dimers are
displaced towards and combine with the rebonded atoms, forming
tetramers T [Fig.~\ref{268-fig}(b)]. The two models in
Figs.~\ref{268-fig}(a) and (b) are nearly degenerate, with a
surface energy difference of $\approx 0.5$ meV/\AA$^2$ relative to
one another. A different number and ordering of the structural
motifs that are present in Fig.~\ref{268-fig}(a) gives rise to a
noticeably larger surface energy. Specifically, the arrangement of
dimers D and honeycombs H [Fig.~\ref{268-fig}(c)] increases the
surface energy by 4 meV/\AA$^2$ relative to the model
\ref{268-fig}(a). An even higher-energy reconstruction made up of
tetramers and rebonded atoms is shown in Fig.~\ref{268-fig}(d).

When the number of atoms in the simulation cell is $n=269$, we
find that the ground state is characterized by the presence of
pentamers, subsurface interstitials, and rebonded atoms [refer to
Fig.~\ref{269-fig}(a)]. Interestingly, this structure is nearly
flat, with the rebonded atoms being only slightly out the plane of
the pentamers. The pentamers are "supported" by six-coordinated
subsurface interstitials. For the Si(337) cell with $n=269$ atoms,
we find a very strong stabilizing effect of the
pentamer-interstitial configuration: the next higher-energy
structure [Fig.~\ref{269-fig}(b)] is characterized by adatoms and
rebonded atoms, with a surface energy that is $\sim$5 meV/\AA$^2$
higher than that of the ground state in Fig.~\ref{269-fig}(a). We
note that subsurface interstitials have long been reported in the
literature to have a stabilizing effect for the reconstruction of
Si(113) \cite{113dabrowski}. Next, we compare the models described
above with previously published works on the Si(337) structure.

\section{Discussion}

To our knowledge, the first model of the Si(337) surface was
proposed by Baski, Erwin and Whitman in their original work on the
structure of Si(5 5 12) \cite{5512science}. The model contained
surface tetramers and 6-member ring $\pi$-chains, and was similar
to the structures depicted in Fig.~\ref{266-fig}(e,f). We observed
a strong relaxation of the $\pi$-bonded chain, in which one of the
atoms of each 6-ring protrudes out of the surface and settles
well-above its neighbors. The surface energy of the model
\ref{266-fig}(f) is not particularly favorable, although it is the
lowest of all the models containing 6-ring $\pi$-chains that we
have investigated. Heuristically speaking, if the
under-coordinated atoms of the 6-rings are removed (4 atoms per
$(2\times 1)$ unit cell), then one would obtain the model of Hu
{\em et al.} \cite{337-applSSci}, which is made of dimers and
tetramers. The genetic algorithm with $n=266$ (modulo 4) has also
retrieved this model, which is shown in Fig.~\ref{266-fig}(b)
after DFT relaxation.

Models \ref{266-fig}(b,e,f) have a high density of dangling bonds
(db) per unit area, 10db/$a^2\sqrt{16.75}$. While the dangling
bond density may seem to account for the large surface energy of
these models, a close inspection of Table~\ref{table} shows that,
in fact, there is no clear correlation between the dangling bonds
and the surface energy. For instance, at the same number of
unsatured bonds per area, models \ref{266-fig}(b,e,f) span a 4
meV/\AA$^2$ range in surface energy. Furthermore, if we take the
DFT model \ref{266-fig}(a) as reference, we note that an increase
by one db per unit cell [from \ref{266-fig}(a) to
\ref{266-fig}(c)] results in a surface energy increase of
$\approx$7 meV/\AA$^2$; on other hand, an increase of seven db in
the simulation cell [from \ref{266-fig}(a) to \ref{266-fig}(b)]
increases the surface energy by less than 3 meV/\AA$^2$. Similar
arguments for the lack of correlation between missing bonds and
surface energy can be made by analyzing structures with other
numbers of atoms in Table~\ref{table}. At the optimum number of
atoms ($n=268$), we find that with the same dangling bond density,
models \ref{268-fig}(a,b,c) span a 4 meV/\AA$^2$ surface energy
range: this range is so large that it includes the global minimum
as well as higher surface energy local minima that may not be
observable in usual experiments. These findings are consistent
with the statement that the minimization of surface energy is not
controlled solely by the reduction of the dangling bond density,
but also by the amount of surface stress caused in the process
\cite{5512science}. Given that the heuristic approaches may only
account for the dangling bond density when proposing candidate
reconstructions, the need for robust minimization procedures
(e.g., \cite{ptmc105, ga105}) becomes apparent.

A notable configuration with 4db/$a^2\sqrt{16.75}$ is shown in
Fig.~\ref{266-fig}(c), where the bonds that join the (lower corner
of) honeycombs and the neighbors of the rebonded atoms are
markedly stretched beyond their bulk value. The stretched bonds
[indicated by arrows in Fig.~\ref{266-fig}(c)] cause the surface
energy to become very high, 100.39 meV/\AA$^2$ at the DFT level.
Interestingly, the surface stress (and surface energy) is very
efficiently lowered by adding dimers such that stretched bonds in
\ref{266-fig}(c) are broken, and the dimers "bridge" over as shown
in Fig.~\ref{268-fig}(a). The addition of one dimer per simulation
cell results in the most stable Si(337) reconstruction that we
found, with $\gamma \approx 89$ meV/\AA$^2$. A structure similar
to \ref{268-fig}(a) was previously proposed by Liu and coworkers
\cite{5512-TEM-takeguchi} to explain their HRTEM data. These
authors also proposed a possible configuration for the other unit
of (337) that is part of the large Si(5 5 12) unit cell, which has
essentially the same bonding topology as \ref{268-fig}(b).
Although no atomic scale calculations were performed in Ref.
\cite{5512-TEM-takeguchi}, the two (337) models of Liu {\em et
al.} were in qualitative agreement with more recent DFT results by
Jeong {\em et al.} \cite{5512-Jeong-yr2004}, as well as with STM
and HRTEM images. When performing ab-initio relaxations, we found
that the HOEP models that corresponded most closely to those in
Ref.~\cite{5512-TEM-takeguchi} end up relaxing into the
configurations put forth by Jeong and coworkers
\cite{5512-Jeong-yr2004}. The structures shown in
Fig.~\ref{268-fig}(a,b) allow for (337) nanofacets to intersect
(113) facets without any bond breaking or rebonding. This absence
of facet-edge rebonding, as well as the relatively low energy of
Si(337), give rise to short-range attractive interactions between
steps on miscut Si(113) \cite{steps-on-Si113}, which is consistent
with the experimental observation of (337) step-bunched phases
\cite{337step-Takeguchi}.

\section{Conclusions}
In summary, we have presented an investigation of the
Si(337)-($2\times 1$) reconstructions based on a genetic algorithm
search for the most favorable atomic configuration \cite{ga105,
ga-perspective}. We have coupled the algorithm with the HOEP model
\cite{hoep} for atomic interactions in order to efficiently create
a database of  Si(337) models. Since no empirical interatomic
potential can be expected to be fully transferable to arbitrary
surface orientations, we have relaxed the database reconstructions
using ab initio DFT calculations. The DFT relaxations give a more
accurate idea about the relative stability of the reconstructions
found, and help identify the likely lowest-energy candidates that
are shown in Fig.~\ref{268-fig}(a) and (b). As it turns out, the
two models coincide with those proposed by Jeong and coworkers
\cite{5512-Jeong-yr2004} from STM experiments and density
functional calculations. This finding lends strong support to the
minimization approach taken here and shows that sampling the
configuration space using the HOEP potential \cite{hoep} is a
versatile way to find good candidate reconstructions for
high-index Si surfaces. We hope that the development of algorithms
for surface structure determination \cite{ptmc105, ga105} would
also stimulate further research towards interatomic potentials
with improved transferability, since the use of such potentials
will improve the quality of the model structures found during the
global optimization: in turn, this would increase the range of
materials problems where the search algorithms can be used to gain
knowledge of optimal structure.

\begin{figure}
  \begin{center}
   \includegraphics[width=12.0cm]{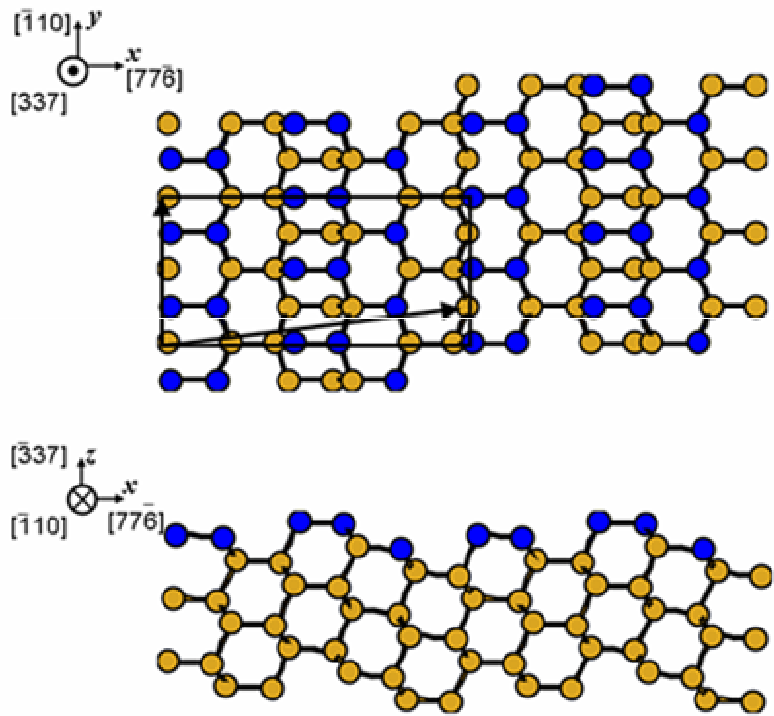}
  \end{center}
  \caption{COLOR ONLINE ONLY. {\bf Chuang, Ciobanu, Wang, Ho}} \label{supercell}
\end{figure}

\begin{figure}
  \begin{center}
   \includegraphics[width=12.0cm]{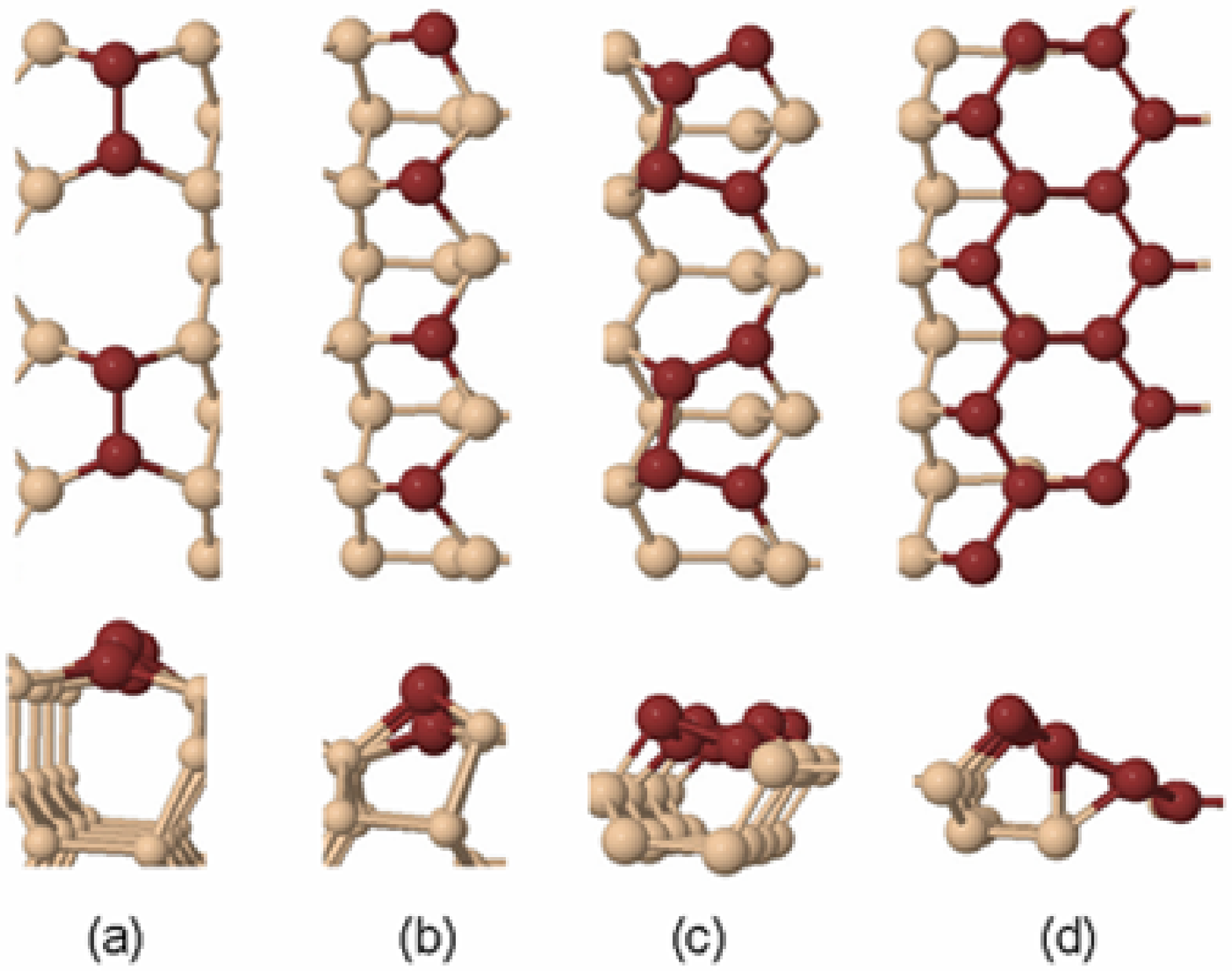}
  \end{center}
  \caption{COLOR ONLINE ONLY. {\bf Chuang, Ciobanu, Wang, Ho}}\label{features}
\end{figure}

\begin{figure}
  \begin{center}
   \includegraphics[width=14cm]{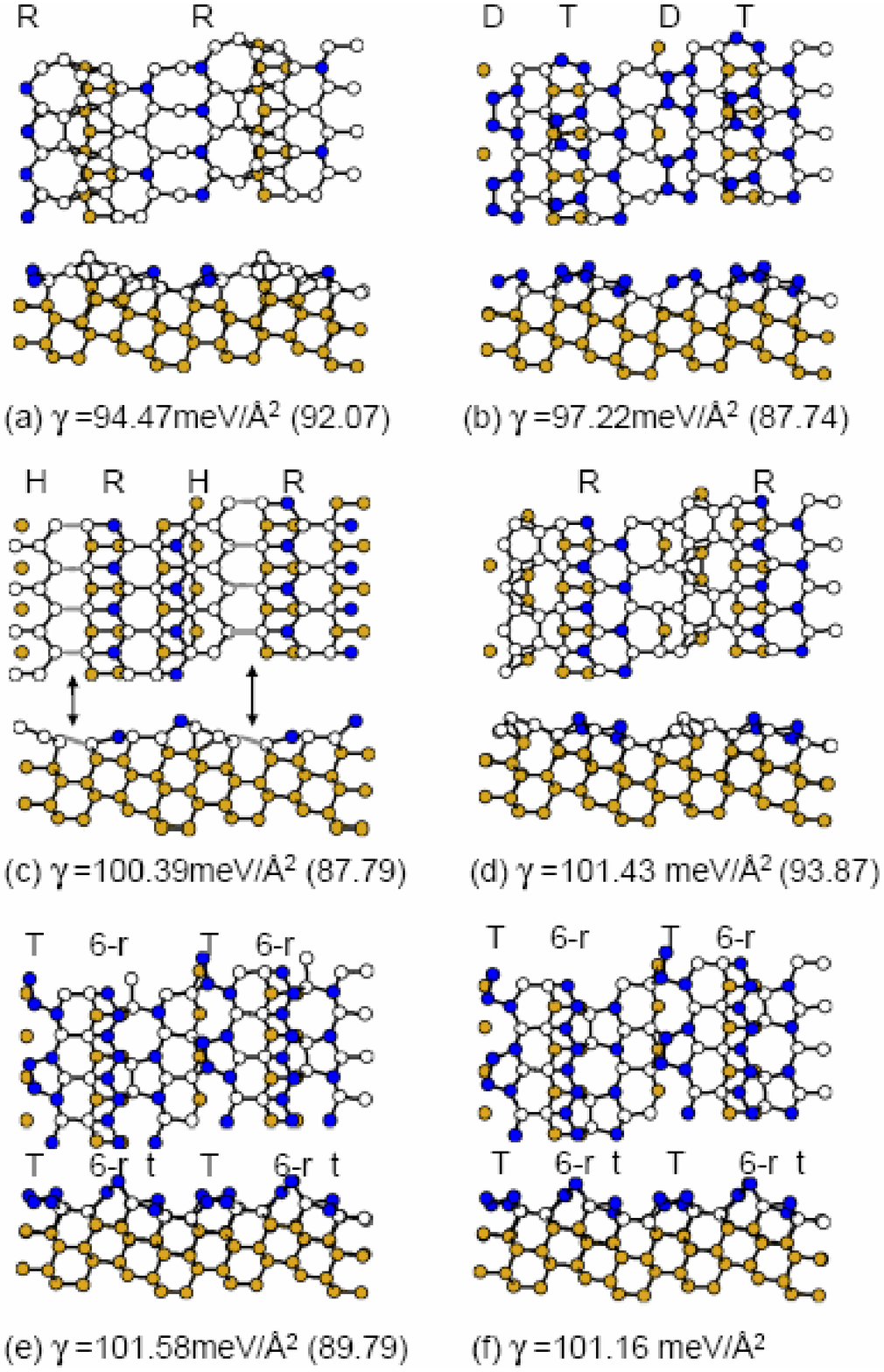}
  \end{center}
  \caption{COLOR ONLINE ONLY. {\bf Chuang, Ciobanu, Wang, Ho}} \label{266-fig}
\end{figure}

\begin{figure}
  \begin{center}
   \includegraphics[width=12cm]{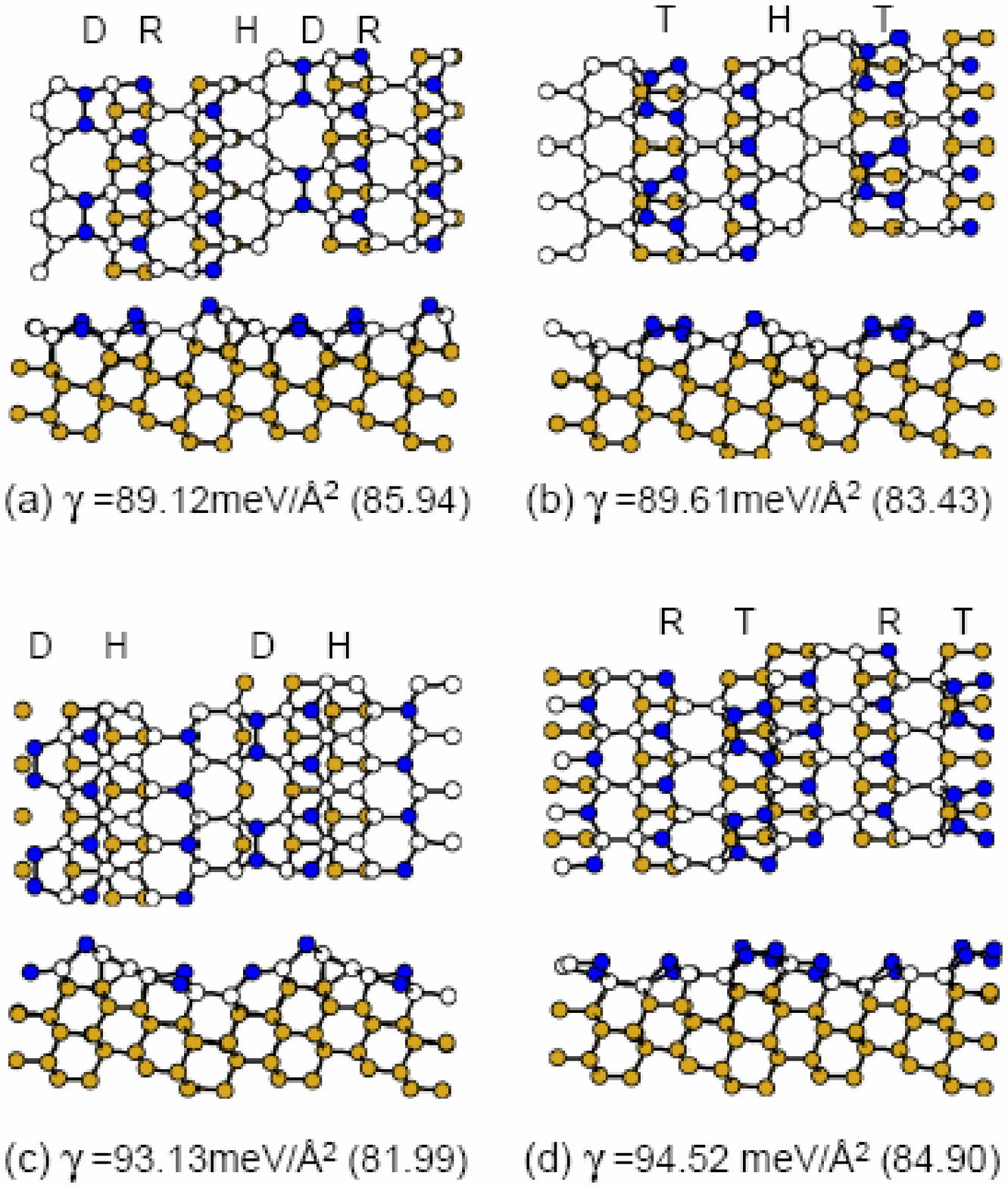}
  \end{center}
  \caption{COLOR ONLINE ONLY. {\bf Chuang, Ciobanu, Wang, Ho}} \label{268-fig}
\end{figure}

\begin{figure}
  \begin{center}
   \includegraphics[width=12cm]{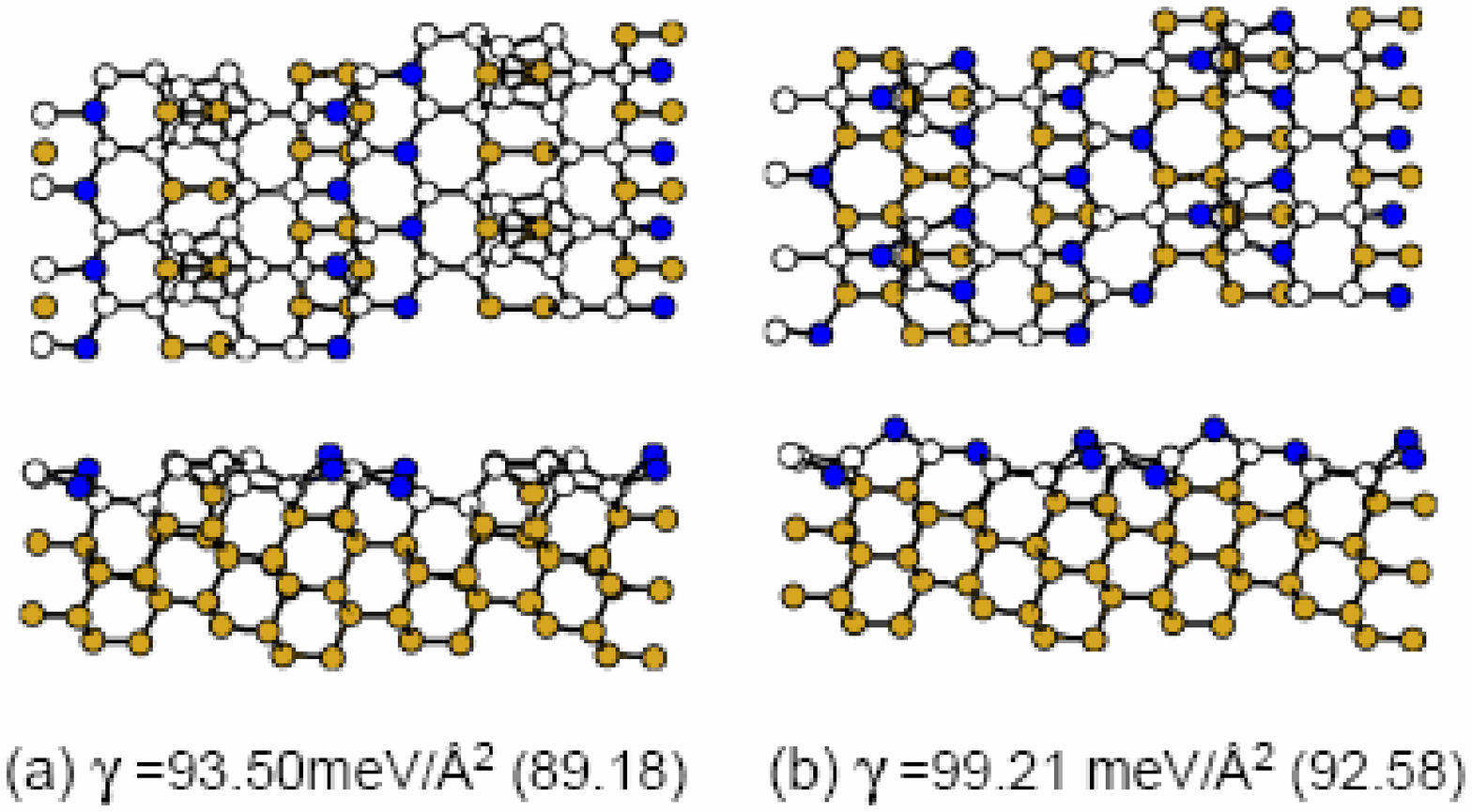}
  \end{center}
  \caption{COLOR ONLINE ONLY. {\bf Chuang, Ciobanu, Wang, Ho}} \label{269-fig}
\end{figure}

The DFT surface energies of the (337) models shown in
Fig.~\ref{268-fig}(a,b) are relatively small, being about 1
meV/\AA$^2$ higher than the surface energy of Si(114) computed
using the same computational parameters Section 2. Upon analyzing
the results of the genetic algorithm optimization, we have seen
that the (337) structures with low surface energies correspond to
certain spatial arrangements of atomic-scale motifs
(Fig.~\ref{features}), rather than to a minimum number of dangling
bonds per unit cell. In experiments, particular conditions can
readily appear such as to favor the stabilization of Si(337) over
Si(5 5 12), as shown for instance in Refs.~\cite{ranke-337-stable,
337Gardeniers1989}. Recent works by Baski and coworkers
\cite{Au-on-Si5512} reveal a controlled way in which Si(337)
facets can be formed --the deposition of gold. At Au coverages of
0.15 ML, the (5 5 12) orientation facets into Si(113) and Si(337),
with the (337) surface unit cells being twice as large as the ones
in Fig.~\ref{supercell} along the $[77{\overline 6}]$ direction.
Since the Au coverages at which Si(337) emerges correspond to only
a couple of Au atoms per each (337) unit, the models presented
here can be used as building blocks for the structure of
Si(337)-Au nanofacets observed in \cite{Au-on-Si5512}, as well as
in other experimental situations.

{\bf Acknowledgments.} CVC thanks J. Dabrowski, S.C. Erwin, S.
Jeong, and M. Takeguchi for correspondence on their model
reconstructions \cite{113dabrowski,
5512science,5512-Jeong-yr2004,5512-TEM-takeguchi}. Ames Laboratory
is operated for the U.S. Department of Energy by Iowa State
University under Contract No. W-7405-Eng-82; this work was
supported by the Director of Energy Research, Office of Basic
Energy Sciences. We gratefully acknowledge grants of supercomputer
time from EMSL at Pacific Northwest National Laboratory, from NCSA
(DMR-050031N), and from NERSC.

\newpage
\renewcommand{\baselinestretch}{1}
\begin{table}
\begin{center}%
\caption{Surface energies of selected Si(337)-$(2\times 1)$
reconstructions, sorted by the number of atoms $n$ in the periodic
cell. The second column shows the number of dangling bonds per
unit area, counted after relaxation with HOEP; the dangling bond
density at the DFT level is shown in parentheses. Columns three
and four list the surface energies given by the HOEP potential
\cite{hoep} and by density functional calculations \cite{pwscf}.}
\label{table} \vspace{1cm}
\begin{tabular}{crccr}
\hline \hline
$n$ &Bond counting              & HOEP              &DFT            & Fig.    \\
    &(db/$a^2\sqrt{16.75})$     & (meV/\AA$^{2}$)   &(meV/\AA$^{2}$)&         \\  \hline

266 &10 (10)                   & 87.37             & 109.04        &                  \\
    &10 (10)                   & 87.74             &  97.22        & \ref{266-fig}(b) \\
    &6  (4)                    & 87.79             & 100.39        & \ref{266-fig}(c) \\
    &10 (10)                   & 89.60             & 102.35        & \\
    &10 (10)                   & 89.79             & 101.58        & \ref{266-fig}(e) \\
    &   (10)                   &  ---              & 101.16        & \ref{266-fig}(f) \\
    &4  (3)                    & 92.07             &  94.47        & \ref{266-fig}(a)  \\
    &4  (4)                    & 93.87             & 101.43        & \ref{266-fig}(d)  \\
\\
267 &8  (8)                    & 92.35             &  99.95        &                   \\
    &8  (8)                    & 92.37             &  96.47        &                   \\
    &10 (8)                    & 92.47             & 107.72        &                   \\
    &9  (8)                    & 99.11             & 101.33        &                   \\
\\
268 &8  (6)                    & 81.99             &  93.13        &\ref{268-fig}(c)  \\
    &8  (8)                    & 83.11             &  95.17        &                   \\
    &8  (6)                    & 83.43             &  89.61        &\ref{268-fig}(b)  \\
    &8  (8)                    & 84.90             &  94.52        &\ref{268-fig}(d)  \\
    &8  (6)                    & 85.47             &  90.19        &                   \\
    &8  (6)                    & 85.94             &  89.12        &\ref{268-fig}(a)  \\
\\
269 &4  (4)                    & 89.18             &  93.50        &\ref{269-fig}(a)  \\
    &6  (6)                    & 92.58             &  99.21        &\ref{269-fig}(b)  \\
\hline\hline
\end{tabular}
\end{center}
\end{table}

{\bf \ \ \ }\newline %
\renewcommand{\baselinestretch}{2}
\newpage
{\bf Figure Captions. All figures COLOR ONLINE ONLY}\newline %
Fig. 1. Top and side view of the unreconstructed Si(337) surface.
The rectangle represents the simulation cell that is subjected to
shifted periodic boundary conditions \cite{farid}, and the solid
arrows are the corresponding periodic vectors; four such periodic
cells are shown in the figure in order to aid the eye with the
shifted boundary conditions. The darker shade marks
the undercoordinated surface atoms. \newline %

Fig. 2. Structural features (top and side views) that can be
present on low-energy Si(337) reconstructions: (a) dimers, (b)
rebonded atoms, (c) tetramers, and (d) honeycombs.
In each case, the atoms that make up the motif are shown in darker shade. \newline %

Fig. 3. Models of Si(337) reconstructions with $n=266$ atoms per
$(2\times 1)$ unit cell, after DFT relaxation (top and side
views). The surface energy computed from first-principles is
indicated for each structure, along with the corresponding value
(in parentheses) determined with the HOEP interaction model
\cite{hoep}. The dark shade marks the undercoordinated atoms. The
four-coordinated atoms that are exposed at the surface are shown
in white. Apart from the relaxations, dimer tilting, and perhaps
the relative phase of dimerization, the structure shown in panel
(b) is the same as that proposed in Ref.~\cite{337-applSSci}, and
reconstructions in panels (e) and (f) are similar to the
model in Ref.~\cite{5512science}. \newline %

Fig. 4. Si(337) reconstructions with $n=268$ atoms per $(2\times
1)$ unit cell, after DFT relaxation (top and side views). The
surface energy computed from first-principles is indicated for
each structure, along with the corresponding value (in
parentheses) determined with the HOEP interaction model
\cite{hoep}. The dark shade marks the undercoordinated atoms,
while the four-coordinated atoms that are exposed at the surface
are shown in white. The lowest-energy Si(337)-$(2\times 1)$
reconstructions are shown in panels (a) and (b). These structures
differ in the position of the dimers in the unit cell: the dimers
D can be part of a 7-member ring [side view, panel (a)], or be
part of a tetramer T [panel (b)]. The models shown in panels (a)
and (b) are the same as those reported in
Ref.~\cite{5512-Jeong-yr2004} for the two (337) units that are
part of a (5 5 12) cell. \newline %

Fig. 5. Models of Si(337) reconstructions with $n=269$ atoms per
$(2\times 1)$ unit cell, after DFT relaxation (top and side
views). The surface energy computed from first-principles is
indicated for each structure, along with the corresponding value
(in parentheses) determined with the HOEP interaction model
\cite{hoep}. The dark blue marks the undercoordinated atoms, while
the four-coordinated atoms that are exposed at the surface are
shown in white. The best $n=269$ model [panel(a)], is stabilized
by flat pentamers and subsurface interstitials. \newline %

\newpage

\newpage 

\end{document}